\documentclass[12pt]{article}
\usepackage{amsmath}
\usepackage{amssymb}
\usepackage{epsfig}
\usepackage{cite}
\usepackage{color,colordvi}

\begin{document}
\vspace{0.01cm}
\begin{center}
{\Large\bf  Black Hole's Information Group } 

\end{center}

\vspace{0.1cm}
%\end{center}

\begin{center}

{\bf Gia Dvali}$^{a,b,c}$ and {\bf Cesar Gomez}$^{a,e}$\footnote{cesar.gomez@uam.es}

\vspace{.6truecm}

%\vspace{.2truecm}

{\em $^a$Arnold Sommerfeld Center for Theoretical Physics\\
Department f\"ur Physik, Ludwig-Maximilians-Universit\"at M\"unchen\\
Theresienstr.~37, 80333 M\"unchen, Germany}

%\vspace{.2truecm}

{\em $^b$Max-Planck-Institut f\"ur Physik\\
F\"ohringer Ring 6, 80805 M\"unchen, Germany}

%{\em $^c$CERN,
%Theory Department\\
%1211 Geneva 23, Switzerland}
%

%\vspace{.2truecm}

{\em $^c$Center for Cosmology and Particle Physics\\
Department of Physics, New York University\\
4 Washington Place, New York, NY 10003, USA}

{\em $^e$
Instituto de F\'{\i}sica Te\'orica UAM-CSIC, C-XVI \\
Universidad Aut\'onoma de Madrid,
Cantoblanco, 28049 Madrid, Spain}\\

\end{center}

%\vspace{0.5cm}

\begin{abstract}
\noindent  
 
{\small 
 We suggest a group-theoretic approach to black holes, which is remotely analogous to the eightfold-way 
 for mesons. 
As the black hole symmetry group we single out the group $SO(2N+1)$ with $N$ the black hole entropy. The Hilbert space is identified with the spinor irrep of $SO(2N+1)$. Evaporation processes of $m$ quanta are associated to the breaking  $SO(2N+1) \rightarrow SO(2(N-m)+1)\otimes SO(2m)$. Under these assumptions we get a group-theoretic understanding of the evaporation process and of some typical time scales of black holes, such as Page's and scrambling times. We also discuss from the group theory point of view the mechanism of generation of entanglement both between the black hole and the radiated quanta as well as among the black hole constituents themselves.}

\end{abstract}

\thispagestyle{empty}
\clearpage

\section{Introduction}

 It is evident that a description in terms of purely geometric entities cannot capture some of the most important black hole properties,  such as, for example,  entropy and/or information processing. 
 The understanding of such properties requires a microscopic description that resolves black hole's quantum constituency.  
  
    We believe that the quantum constituency of  macroscopic black holes must become apparent 
 already at distances comparable to their classical radius, $R$.   This constituency must be 
 largely insensitive to the particular form of UV-completion of gravity at microscopic distances, e.g., such as 
 the Planck length, $L_P$.   

  Some time ago \cite{Nportrait} we have outlined how this quantum picture comes about.   A black hole of a classical radius $R$,  in reality represents a Bose-Einstein condensate 
  (BEC) of soft (wavelength $\sim \,  R$)  gravitons stuck at the critical point of a {\it quantum phase transition}.   For the system of gravitons of wavelength R, the  quantum criticality is reached when the  occupation number of gravitons is inverse of their gravitational coupling, 
  \begin{equation}
      N \, = \, \alpha^{-1}  \, \equiv \, (R/ L_P)^2  \, . 
  \label{alpha}
 \end{equation} 
 One can immediately notice, that the occupation number at the critical point scales as area, as opposed to the volume.   This is the key to understanding the scaling of the black hole entropy in our picture. 
 
 The advantage of this microscopic picture is that it allows to address the questions which in standard 
 semi-classical treatment cannot even be consistently posed.  For example,  it allows to monitor the  
 underlying mechanism for information scrambling.  It was argued some time ago  \cite{scrambling} that the  black holes must scramble information within the time that scales as log of the area, but without having a microscopic framework it was impossible to either verify this claim or to understand the underlying quantum mechanism behind it.  It was shown recently \cite{us} that Bose-Einstein picture of black holes  reveals 
 the key mechanism behind scrambling in form of a quantum break time of an unstable condensate, and  predicts the log$N$ scrambling time  in full accordance with \cite{scrambling}. 
% This makes us think that we are on a right track towards understanding the quantum portrait of black holes.  
 
  While the studies towards understanding various aspects of this proposal are ongoing, in the present note we shall offer a {\it symmetry-group} approach to black holes.  
  
    From our quantum portrait we shall adopt the fundamental concept that 
   a macroscopic black hole is a composite system of  $N$ quantum constituents
   and that the collective effects of the constituents, such as the appearance of gapless
   Bogoliubov modes,  are maximally important.

   Then,  we shall try to derive quantum properties of the black holes, by postulating a simple 
  symmetry group structure for its quantum constituents.  In this respect,  our approach is 
  analogous to the  "eightfold way" of mesons in which their properties are derived from a symmetry structure of the constituent quarks. In essence we postulate that the black hole dynamics is subject to a symmetry group, which we denote as
      the {\it BH-information group}. This will allow us to derive some generic aspects of black hole evaporation dynamics in pure group theoretical terms.

The rest of our discussion will be independent of our BEC portrait, which we use simply 
      as evidence for black hole compositeness. 
     The reader can fully abstract from this underlying picture and take our symmetry approach as an effective guiding principle, much in the same way as one can abstract from QCD dynamics and 
     try to understand properties of hadrons from symmetry principles of quarks.

\section{The BH Symmetry Group}
	From now on we shall reduce ourselves to Schwarzschild-like black holes that can be uniquely characterized in terms of the value $N$ of the Bekenstein-Hawking entropy  \cite{bekenstein, hawking} . Our main postulate will be to identify the Hilbert space of a black hole of entropy $N$ with the unique fundamental spinor irrep of $SO(2N+1)$. This irrep that we shall denote by $[N]$ has dimension $2^N$ and therefore we can define, as it is customary, the black hole entropy as the log of the dimension of the Hilbert space of states. Thus our first postulate can be summarized by the following correspondence
\begin{equation}
BH(N) \rightarrow [N]
\end{equation}
Correspondingly we shall identify $SO(2N+1)$ as the BH-symmetry group. Before going on, let us recall few basic facts about the irreps of $SO(2N+1)$. For the group $SO(2N)$ we have two fundamental spinor irreps that differ by the corresponding chirality. These irreps that we shall denote $[N]_{+}$ and $[N]_{-}$ have dimension $2^{N-1}$. The irrep $[N]$ of $SO(2N+1)$ is simply the direct sum of these two chiral spinor irreps:
\begin{equation}
[N] = [N]_{+} \oplus [N]_{-}
\end{equation}
The simplest way to visualize the chiral spinor irreps of $SO(2N)$ is as a fermionic Fock space. Concisely we define the algebra of $N$ creation $a^i$ annihilation $a_i$ operators satisfying:
\begin{equation}\label{three}
\{a_r, a^s\} = \delta_{r}^{s} I
\end{equation}
with $\{a_r,a_s\} =\{a^r,a^s\}=0$ and $I$ the unit operator. The two chiral irreps are spanned by Fock space vectors
$\prod a^i |0 \rangle $ with  even or odd value for the number operator respectively. Both subspaces have dimension $2^{N-1}$ and together they span the whole fundamental spinor irrep of $SO(2N+1)$.

In an obvious holographic interpretation \cite{hologram} we can think of the $N$ operators $a^i$ as $N$ different letters and the different vectors spanding the irrep $[N]$ as the whole set of messages we can write in terms of the $N$ BH holographic bits. 

Already at this level of the discussion we can identify the symmetry breaking pattern of the black hole evaporation process. Indeed, in one evaporation step, irrespectively what can be the underlying dynamical mechanism, we expect to go from a black hole of entropy $N$ to one of entropy $N-1$. From the point of view of the BH symmetry group this means that we should {\it break}
\begin{equation}
SO(2N+1) \rightarrow SO(2(N-1)+1)
\end{equation}
and generically in $m$ evaporation steps,  $SO(2N+1) \rightarrow SO(2(N-m)+1)$. Our next task will be to identify the group-theoretic meaning of this symmetry breakdown induced by the evaporation process.

\subsection{The group theory meaning of BH-evaporation}
In order to fix ideas, let us start with a BH of entropy $N$ i.e.,  with the irrep $[N]$ of $SO(2N+1)$. We shall model the evaporation process in three steps.

{\bf Step 1.} First we use the freedom to decompose the irrep $[N]$ into the two chiral irreps $[N]_{+}$ and $[N]_{-}$ of $SO(2N)$. Already at this level each of the chiral irreps has the appropiate dimension $2^{N-1}$ to account for the entropy of the BH after the emission of {\it one} quantum, i.e.,  $N-1$. However, if we would simply identify the BH -- after one evaporation step -- with this chiral irreps we would have to change the BH symmetry group for the new BH to be $SO(2N)$ as well as to assign to the BH a fictitious chirality. The way to avoid these undesired consequences leads us to the second step of the evaporation process.

{\bf Step 2.} Since the BH symmetry group after one evaporation step is $SO(2(N-1)+1)$, what we should do is to map the two chiral irreps we have obtained in step 1 above into the {\it unique} spinor irrep $[N-1]$ of the new BH symmetry group $SO(2(N-1)+1)$. In other words, after one evaporation step the two chiral irreps $[N]_{+}$ and $[N]_{-}$ are identified with the unique spinor irrep $[N-1]$ of the new BH symmetry group. In the next step we need to identify what happens with the two chiral labels we are missing by this identification.

{\bf Step 3} Since we are keeping ourselves in the full Hilbert space of the original BH we can identify the part of the Hilbert space that corresponds to the BH after evaporation as well as the part of the Hilbert space of the emitted quanta. The BH Hilbert space is the irrep $[N-1]$ we have defined in step 2 above. The two chiral labels define the two possible states of the emitted quantum that we can interpret as the two chiral spinor irreps of $SO(2)$. In other words the chirality we have used in the intermediate step 2 is {\it transmuted} into the labels of the  Hilbert space of the radiated quanta.

After completing the description of the group-theoretic steps of the BH evaporation we
are able to guess the symmetry-breaking pattern of the BH evaporation process, namely:
\begin{equation}
SO(2N+1) \rightarrow SO(2(N-1)+1) \otimes SO(2) \, .
\end{equation}
Obviously, the process can be continued to successive evaporation steps. For instance, if we consider two emitted quanta we will get the irrep $[N-2]$ for the BH as well as a Hilbert space for the two quanta of dimension four corresponding to the two spinor irreps of $SO(4)$. Thus, after $m$ evaporation steps, the symmetry breaking pattern is
\begin{equation}
SO(2N+1) \rightarrow SO(2(N-m)+1) \otimes SO(2m) \, .
\end{equation}
Note that the sub-algebra $SO(2(N-m)+1) \otimes SO(2m)$ is the maximal regular sub-algebra, i.e.,  the maximal sub-algebra having the same Cartan algebra as $SO(2N+1)$. In pictorial terms each evaporation step can be represented as removing a node in the extended Dynkin diagram.

The previous group-theoretic picture gives us a natural prescription for writing the typical quantum state after $m$ evaporation steps, namely
\begin{equation}\label{one}
|\Psi \rangle \,  =  \, \sum_{i=1}^{2^m} |\psi_{i} \rangle \otimes |i \rangle  \, ,
\end{equation}
with
\begin{equation}
|\psi_{i} \rangle  \,  \in  \,   [N-m]
\end{equation}
for $[N-m]$ the spinor irrep of $SO(2(N-m)+1)$. More concretely, if we start with a state $|\psi\rangle  \in [N]$ for the initial black hole of entropy $N$,  then -- after one evaporation step -- we shall generically get a state $|\psi_+ \rangle  \otimes |+ \rangle \,  + \, |\psi_-\rangle \otimes |- \rangle $ with $|\psi_{\pm} \rangle  \in [N-1]$. Since $[N-1]=[N-1]_{+}\oplus [N-1]_{-}$,  we can think as the most natural possibility that $|\psi_{\pm}\rangle \in [N-1]_{\pm}$. Subsequently,  the process can be repeated in the next step for each state $|\psi_{\pm} \rangle$ leading to four states $|\psi_{\pm;\pm} \rangle  \in [N-2]$, and so on. 

\subsection{Group theory approach to BH time scales}
It is pretty obvious that the state (\ref{one}) represents entanglement between the BH state and the radiated quanta. We can now ask ourselves when this entanglement becomes maximal. From the group theory perspective the answer is very simple, namely it will become maximal whenever the dimension of the BH irrep $[N-m]$ is equal to the dimension of the Hilbert space of the radiated quanta, i.e.,
\begin{equation}
2^m \, = \,  2^{N-m} \, ,
\end{equation}
which gives $m\, = \, m_{Page} \, = \, N/2$,  i.e., Page's time \cite{Page}. Obviously, at this point the number of black hole states entering into (\ref{one}) is exactly equal to the dimension of the corresponding black hole Hilbert space.

What about the group theory meaning of scrambling time? It is easy to observe that in $m$ evaporation steps we create $2^m$ {\it chirality labels}. Thus, in order to create order-$N$ chirality labels we need a number of steps scaling with $N$ as ${\rm log}  N \,$,  i.e.,  as the scrambling time. It is amusing to observe that this is the time needed to get a Hilbert space for the radiated quanta with dimension of the order of the dimension of the fundamental (not spinor) irrep of the BH symmetry group. However, this is not telling us too much about the meaning of scrambling time. Indeed, its meaning can be unveiled only when we track the time evolution of the BH state itself. In the next section we shall address this issue.

\section{Time evolution of the BH state}
Our task in this section is to try to figure out how the BH state evolves along the evaporation process using as a guiding principle the group-theoretic approach we have developed in the previous sections. In more concrete terms we start with a particular BH state $|BH(N) \rangle  \in [N]$ and we track the evolution of this state into the BH state $|BH(N-m) \rangle  \in [N-m]$. 

Using again the decomposition $[N]= [N]_{+} \oplus [N]_{-}$ we can generically represent $|BH(N) \rangle $ as
\begin{equation}\label{two}
|BH(N) \rangle \,  = \,  |W_{+} \rangle \,  + \,  |W_{-} \rangle\, , 
\end{equation}
for $|W_{\pm} \rangle \,  \in \, [N]_{\pm}$ respectively. Now the BH state after one emission will admit a similar decomposition but in terms of some new states $|w_{\pm} \rangle \,   \in  \, [N-1]_{\pm}$. A priori, the BH of initial entropy $N$ can be in a state $[N]$ with a well-defined chirality. However, since the BH symmetry group is $SO(2N+1)$, generically the evolution of the state along the evaporation process will not preserve chirality. Thus, even if we start with one state of definite chirality, we shall generically expect to get after one evaporation  step a superposition state $|w_+\rangle  \, + \, |w_- \rangle $. 

The generators of $SO(2N+1)$, which are not in $SO(2N)$,  i.e.,  the ones intertwining different chiralities, are
\begin{equation}
J_{2r} \, = \, - \, \frac{1}{2}i \, (a_r \, - \, a^r)
\end{equation}
and
\begin{equation}
J_{2r-1} \,  =\,  -\, \frac{1}{2}\, (a_r+a^r)
\end{equation}
with $r=1,2...N$ and with $a_r$ and $a^r$ the algebra operators defined in (\ref{three}). They define a Clifford algebra $\it{C(N)}$ of gamma matrices; $\gamma_k \, = \, 2 J_k$,
\begin{equation}
\{\gamma_k,\gamma_l\} \, =\,  \delta_{k,l}I \, .
\end{equation}

In order to figure out the evolution of the black hole state along the evaporation process, let us write the state (\ref{one}) after one evaporation step as
$|\psi_+ \rangle \otimes|+ \rangle \,  + \, J|\psi_+ \rangle \otimes |- \rangle $
with $J\in \it{C(N-1)}$. Thus,
a simple way to imagine the evolution along the evaporation process is like a random path of actions  on the initial state. In order to fix ideas, let us consider a basis vector $ |\epsilon_1...\epsilon_N \rangle \,$,  where $\epsilon_r=\pm$  are the eigenvalues of the Cartan sub-algebra generators 
$\sigma_{r} \, \equiv \, - i \gamma_{2r-1}\gamma_{2r} \, = \, (a^ra_r \, - \, a_ra^r)$.
Let us represent it as $ |\epsilon_1...\epsilon_{N-1} \rangle \, \otimes |\epsilon_N \rangle $. Let us now act on this state with the operator
\begin{equation}\label{four}
U_r = (\alpha \, I \otimes I\,  + \,  \beta \,  (a_r \, + \, a^r) \otimes (a_N \, + \, a^N)) \, , 
\end{equation}
where $\alpha, \beta$ are parameters and  the index  $r \neq N$ is  otherwise chosen randomly. 
%$\in \it{C(N-1)}$ with $i$ chosen randomly and where 
This operation creates an entangled superposition between the emitted state $|\epsilon_N \rangle$ 
and the remaining $(N-1)$-particle basis vector  $ |\epsilon_1...\epsilon_{N-1} \rangle$. 

For the next step of the evaporation we perform the same operation over $(N-1)$-particle basis vectors
of the BH state.   
In this way, we act along the evaporation of $m$ quanta using a random path $(r_1,r_2...r_m)$. 

 The physical meaning of the above sequence is easy to understand by noticing that 
 the operators  $(a_r \, + \, a^r) \otimes (a_N \, + \, a^N)$  are the broken generators of 
 $S0(2N +1)$ that act non-trivially 
 on $S0(2(N-1) +1)$ and $SO(2)$ spinor spaces.  In the language of spontaneous symmetry breaking they correspond to Nambu-Goldstone bosons of the broken information group.  
  Thus, we can say that the  entanglement is generated due to excitement of Goldstone bosons of broken 
  information group in every act of emission.

From the previous construction we can derive two important results regarding the creation of entanglement at the level of the BH state itself. It is important not to confuse this entanglement for the BH state as representing a composite system with the entanglement between the BH and the radiated quanta. In the case of a random path of applications of operators of type (\ref{four}) it is clear that we create entanglement with each step. Of course, the entanglement will be maximal when the number of different states entering into the superposition at the end of $m$ steps is equal to the dimension of $[N-m]$. Since in each action of (\ref{four}) we generically create a superposition of two states the number of states entering into the superposition after $m$ steps is $2^m$. So maximal entanglement requires $2^{N-m}\, = \, 2^m$, which again reproduces Page's time,  i.e.,  $m \, = \, N/2$. 

With respect to scrambling time we observe that it is the time needed to create in a random path, i.e.,  in a path where the same operator never repeats, a superposition of $N$ states. Indeed, the time required to get such a superposition is determined by $2^m \, = \, N$,  i.e., 
\begin{equation}
m_{scrambling} \, = \,   {\rm log}  N \, .
\end{equation}
 Generically this state will be one-particle entangled. Indeed, for a generic non-entangled initial  state, such as $ |\epsilon_1...\epsilon_N \rangle \,$,   after a random path of $m\, = \, {\rm log}  N \,$ steps we shall get a superposition of $N$-states where 
 none of the eigenvalues $\epsilon_r$ will have the same value throughout the final state \footnote{
 Notice that  probability  of repetition of a same generator in such a random sequence  of 
 ${\rm log}  N$-steps is suppressed at least as $\sim \, 1/N $. }.
  This is enough to guarantee one-particle entanglement.

\section{Final Comment}

From the group theory perspective developed in this paper at each step in the evaporation process the black hole symmetry is reduced to a maximal regular sub-algebra governing the radiated quanta as well as the remaining black hole.
As described above,  one of the main clues -- of this group-theoretic setup -- for the understanding of the
black hole evaporation is to associate the source of entanglement between the 
radiated quanta and the black hole to the generators of the black hole symmetry group that are "broken" in the evaporation process. 
Thus, the entanglement is generated by exciting the Goldstone bosons of the spontaneously broken 
information group. 

 The interesting question is how much of the underlying dynamics is captured by this symmetry picture. 
 This question is only possible to answer within a microscopic theory.  The black hole's 
 quantum portrait \cite{Nportrait}  in form of a critical Bose-Einstein condensate provides such an explicit framework.   In this picture the holographic degrees of freedom become explicit and represent 
 Bogoliubov modes of the critical condensate.  The generation of entanglement and scrambling is directly related to the quantum instability that reduces $N$ as well as with the huge density of states near the critical point \cite{us}.  Thus,  the Goldstone modes of the broken information group would be naturally  identified with the collective modes of the system that connects these degenerate states.  
 
     As a final comment, it would be interesting to explore the role of other possible groups. 
  The dual ( in Langlands $S$-duality sense ) of $SO(2N+1)$ is the group $Sp(N)$ \cite {Witten}. 
  A natural question then would be to explore the role of this "dual" description.  
An amusing possibility would be to associate this couple of dual groups with the two possible versions of the black hole, namely as described by exterior and inside observers.

 \section*{Acknowledgements}

We would like to thank  Sumit Das,  Daniel Flassig,  Alex Pritzel and Nico Wintergerst for discussions.  
The work of G.D. was supported in part by Humboldt Foundation under Alexander von Humboldt Professorship,  by European Commission  under 
the ERC advanced grant 226371,   by TRR 33 \textquotedblleft The Dark
Universe\textquotedblright\   and  by the NSF grant PHY-0758032. 
The work of C.G. was supported in part by Humboldt Foundation and by Grants: FPA 2009-07908, CPAN (CSD2007-00042) HEPHACOS P-ESP00346 and SEV-2012-0249.

\end{document}